\documentclass[onecollarge,natbib]{svjour2}
\bibpunct{[}{]}{,}{n}{}{,} 
\smartqed  

\usepackage{graphicx}
\usepackage{epsfig}
\journalname{Few-Body Systems (APFB2011)}
\begin{document}

\title{\boldmath
The baryon number two system in the Chiral Soliton Model
\thanks{This work  has been supported  by HadronPhysics2,  MICINN (Spain) grant  FPA2010-21750-C02-01,   
AIC10-D-000621 and by GVPrometeo2009/129. }
}

\author{Valentina  Mantovani-Sarti \and Alessandro  Drago \and Vicente Vento  \and Byung-Yoon Park
        }


\institute{V. Mantovani-Sarti and A. Drago \at
              Dipartimento di Fisica, Universit\'a di Ferrara and INFN sezione di Ferrara, 44100 Ferrara, Italy \\
              \email{smantovani@fe.infn.it, drago@fe.infn.it}           
           \and
           V. Vento \at
              Departamento de F\'{\i}sica Te\'orica and IFIC,
Universidad de Valencia - CSIC 
E-46100 Burjassot (Valencia), Spain \\
              \email{vicente.vento@uv.es}  \\
           \and
        B.-Y. Park \at Department of Physics, Chungnam National University, Daejon 305-764, Korea\\
              \email{bypark@cnu.ac.kr}\\
}

\date{Received: date / Accepted: date}

\maketitle

\begin{abstract}
We study the interaction between two $B=1$ states in a Chiral Soliton Model 
where  baryons are described as non-topological solitons. 
By using the hedgehog solution for the $B=1$ states we construct  three possible $B=2$ configurations 
to analyze the role of the relative orientation of the hedgehog quills in the dynamics. 
The strong dependence of the intersoliton interaction on these relative orientations reveals that studies of dense hadronic matter using this model should take into account their implications.

\end{abstract}

\vskip 1.0cm

One  important issue in the hadron physics is to understand the properties of hadronic matter under extreme conditions,  e.g., at high temperature as in relativistic heavy-ion physics and/or at high density as in compact stars. The phase diagram of hadronic matter turns out richer than what has been predicted by perturbative Quantum Chromodynamics (QCD)~\cite{McLerran:2010zz}. Because of the well-known ``sign problem" in Lattice QCD \cite{Karsch:2007dt}  one cannot explore the whole phase space by the direct calculations  from the fundamental theory with  quark and gluon degrees of freedom.  Thus, we must rely on effective field theories which are defined in terms of hadronic fields \cite{Weinberg:1978kz} .

The Skyrme model~\cite{Skyrme:1962vh} is an effective low energy meson theory rooted in large 
$N_c$ QCD, where a baryon is described by a topological soliton ~\cite{Witten:1983tx}.  
This model  has been shown to be successful in describing not only  single baryons but even
heavy baryons and  multi-baryon systems \cite{Brown:2010br}. 
Furthermore, this model has been applied  to  study dense and hot baryonic matter and the consequent modifications of meson properties in such a medium have been obtained \cite{Lee:2003aq,Park:2008xm}.
Since this model does not contain explicit quark and gluon degrees of freedom, the main interest of these studies has been the phase transition associated with  the chiral symmetry restoration.

Another description which has been applied to the study of hadronic properties is the non topological soliton model introduced by Friedberg and Lee \cite{Friedberg:1976eg}.
In this description both quark fields and meson fields are used to obtain the properties of hadrons, although the color degrees of freedom remain hidden. The model has also been  applied to the study of hadronic matter and the chiral phase transition by using a single Wigner-Seitz cell \cite{Hahn:1987xr}.  This formalism, very different from the one used in similar studies in the Skyrme model, disregards some of the features associated with the long range interaction, which play an essential role in describing the phase transition in the Skyrme model. 

We present here our study of  the two body dynamics in the Chiral Soliton Model,
 focusing on the  properties arising from the soliton (non topological) structure of the theory \cite{smantovani}. The understanding of the two body forces, will allow us to describe hadronic matter taking into account the long range effects of the intersoliton dynamics. The new feature in the present approach, in comparison with the Skyrme model, is the description of the quark degrees of freedom.

We use for our analysis the simple chiral quark model studied in ref.\cite{DV11a}.  
The model Lagrangian density reads 
\begin{equation}
\mathcal L = \bar{\psi} (i \gamma^\mu \partial_\mu - g ( \sigma + i \vec{\tau}\cdot \vec{\pi} \gamma_5 ))\psi + 
\frac{1}{2} \left(\partial_\mu \sigma \partial^\mu \sigma +  \partial_\mu \vec{\pi} \cdot \partial^\mu \vec{\pi} \right)  - V(\sigma, \vec{\pi}),
\end{equation}
where $\sigma$ represents a scalar isosinglet meson, $\vec{\pi}$ are isotriplet pion fields and $\psi$ describes the isodoublet ($u$ and $d$ ) quark fields. $V(\sigma, \vec{\pi})$ is a potential energy  that should provide symmetry breaking and we use the model of ref.\cite{DV11a}. This Lagrangian is a generalization of the Friedberg-Lee model which implements the appropriate realization of chiral symmetry.
By choosing the vacuum at $(\sigma_0=f_\pi, \vec{0})$, the model Lagrangian describes mesons and quarks with masses $m_\sigma$, $m_\pi$ and $m_q = gf_\pi$, respectively. 
For these parameters, we  take the following values $m_\sigma = 550$ MeV, $m_\pi = 138$ MeV, $g=5$ and $f_\pi=93$ MeV.  

The self-consistent  $B=1$ solution using the hedgehog Ansatz is given by,
$$ \sigma_{B=1} (\vec{r}) = \sigma (r) \;\; \; ,     \;\; \;  \vec{\pi}_{B=1} (\vec{r})= \pi(r) \hat{r} $$
\begin{equation}
\psi_{B=1} (\vec{r}) = \frac{1}{\sqrt{4 \pi}}
\left(\begin{array}{c} 
u(r) \\ 
i \vec{\sigma} \cdot \hat{r} v(r)
\end{array} 
\right)
 \frac{1}{2} (| u \downarrow \rangle - | d \uparrow \rangle).
 \end{equation}
The solution is stabilized, {\em not} by a topological constraint, {\em but} by the energy which becomes  lower than  three free quark masses. The solution for the the sigma field developes a bag-like spatial structure where the quark fields become localized. Making such bag-like structure costs more than 800 MeV in meson field energy,  which  is compensated by the binding energies of the quarks. In total, the whole system has a binding energy of about 400 MeV. The rms radius of the baryon number distribution is about 0.7 fm.

In order to study the soliton-soliton interaction, we need to construct a $B=2$ system where two solitons are separated by some distance. Suppose that we have two solitons whose centers are at $\vec{r}_1$ and $\vec{r}_2$. 
When two solitons are sufficiently far apart, the ``product Ansatz" from the Skyrme model is useful  to construct the two-soliton system \cite{Lee:2003aq}.  In this scheme the $\sigma$ and $\vec\pi$ configuration for the $B=2$ system can be approximated  by
\begin{eqnarray}
&&\left(\frac{\sigma_{B=2} (\vec{r}) + i \vec{\tau} \cdot \vec{\pi}_{B=2} (\vec{r})}{f_\pi} \right) 
\nonumber \\
&& \hskip 2em   = \left(\frac{\sigma_{B=1} (\vec{r} - \vec{r}_1) + i \vec{\tau} \cdot \vec{\pi} _{B=1} (\vec{r} - \vec{r}_1)}{f_\pi}\right)
  C \left(\frac{ \sigma_{B=1}  (\vec{r} - \vec{r}_2) + i \vec{\tau} \cdot \vec{\pi} _{B=1} (\vec{r} - \vec{r}_2) }{f_\pi} \right) C^\dagger,
\label{ProdAnsatz}\end{eqnarray}
where $f_\pi$ is used to  have the proper dimensions. As was done  in the Skyrme case,
a constant $SU(2)$ matrix $C$ can be inserted to introduce the relative orientation of the second soliton with respect to the first one in isospin space. We  choose $\vec{r}_{1,2} = \pm (d/2) \hat{z}$ and consider three different configurations
\begin{itemize}
\item Configuration A : $C$=1 ,  i.e., two unrotated hedgehog solitons,
\item Configuration B : $C = e^{i \tau_z \pi/2} = i \tau_z$, i.e. , the second soliton is rotated by an angle $\pi$ about the axis that is parallel to the line joining two centers, 
\item Configuration C : $C = e^{i \tau_x \pi/2} = i \tau_x$, i.e. , the second soliton is rotated by an angle $\pi$ about the axis that is perpendicular  to the line joining two centers.
\end{itemize}

\begin{figure}
\begin{center}
 \includegraphics[scale=0.7]{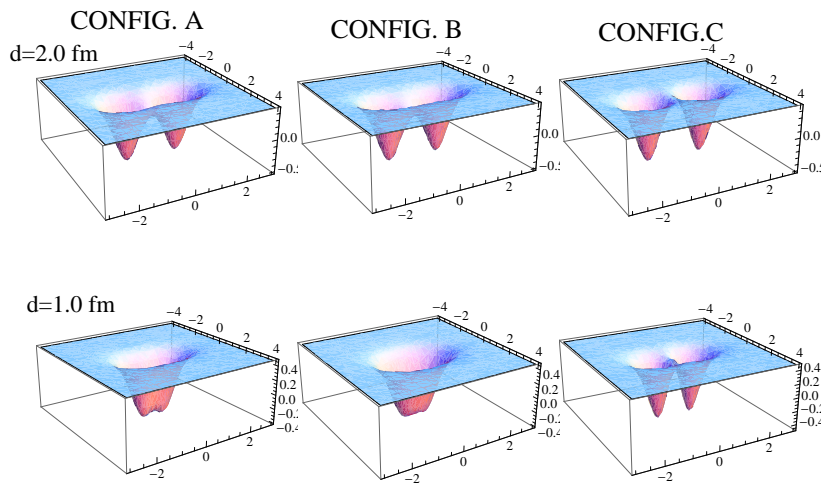}
\caption{The structure of the $\sigma$ field for B=2, $\sigma_{B=2}$, for the three configurations studied at large (upper figures) and small (lower figures) separation parameter $d$.}
\label{fig:1}      
\end{center}
\end{figure}

In the Skyrme model, taking the product of two $B=1$ soliton solutions is one of the most convenient ways to 
obtain the $B=2$ intersoliton dynamics \cite{Lee:2003aq}. However, in the ``linear" Chiral Soliton Model, with explicit quark degrees of freedom, since we are not restricted by a topological winding number, the  {\em product} scheme may not be so essential, though it provides some advantages. First of all, it 
makes $\sigma_{B=2}$ and $\vec{\pi}_{B=2}$ naturally satisfy the boundary conditions at infinity; that is,   
\begin{equation}
\sigma(r \rightarrow \infty) \rightarrow f_\pi, 
\hskip 3em \vec{\pi}(r \rightarrow \infty) \rightarrow 0
\end{equation}
without any further artificial construction. 
Secondly, when the separation distance between two solitons is sufficiently large, the two solitons 
will have their own identity. 
We show in Fig.\ref{fig:1} the $\sigma_{B=2}$ field obtained by this Ansatz (\ref{ProdAnsatz}). 
Note that, at large and intermediate separations, the relative distance is a well defined quantity, while at short separations the baryons deform heavily, making complicated overlapping shapes,  and the relative distance cannot be well defined.  The meson cloud appears as a halo around the baryons and deforms  as the baryons approach eachother. This  deformation is the energy density representation of the  pion exchange potential. The strong deformation of the cores is a consequence of the repulsion mediated in reality by multiple meson exchanges.

The zeroth order solution of the quark wavefunctions, since the background field has some kind of reflection symmetry except for an isospin rotation,  can be expressed as the linear combination 
\begin{equation}
 \psi_{B=2}^{\pm} = \frac{1}{\sqrt2} \left( \psi_{L}(\vec{r}) \pm \psi_R (\vec{r}) \right), \mbox{where} \;  \psi_{L}(\vec{r}) = \psi_{B=1}(\vec{r}-\vec{r}_{1}),  \; \mbox{and} \;
\psi_{R}(\vec{r}) = C \psi_{B=1}(\vec{r}-\vec{r}_{2}), 
\end{equation}
and $\psi_{R}(\vec{r})$ is rotated by $C$.

The energy of the $B = 2$ system is then calculated by substituting these approximated solutions for the mesons 
into the corresponding formulas and evaluating the expectation values of the Hamiltonian with respect to the $B=2$ state where three quarks occupy each energy levels described by the wavefunctions $\psi_{\pm}$. 
By subtracting twice the soliton mass from the total energy of the $B=2$ system, we obtain the soliton-soliton interaction energy as a function of the separation distance and the relative orientation. 

\begin{figure}
\begin{center}
 \includegraphics[scale=0.30]{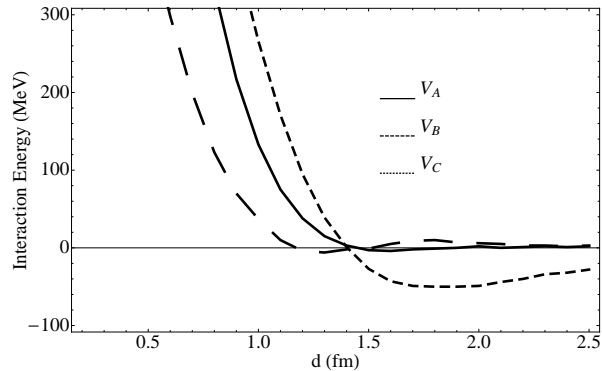}
\caption{Soliton-soliton interaction energy as a function of the separation distance for the three relative orientations mentioned in the text.}
\label{potABC}    
\end{center}  
\end{figure}
In Fig.\ref{potABC} we show the interaction as a function of $d$ for the three different configurations. 
There is  no dramatic  difference in shape between the three cases. The difference is a matter of  detail.
Configuration A and B are repulsive  for all $d$ and B is more repulsive at the large distance. At large distances the 
most stable state is in the C configuration. Thus, at large separation the behavior of the soliton interaction 
seems to be similar to that obtained in the Skyrme model. This result is rather predictable since when two solitons are far apart, the interaction between them is mainly through the  meson exchange. 

As the baryons get closer to each other there is a  transition to the B
configuration, just at the point where $V_C$ rises. Configuration B seems  to be the lowest energy state only where the interaction is repulsive. It seems that the quark fields plays a role at short separation. However, our first order calculations, without any modifications on the quark wavefunction, are too primitive to draw any conclusion for the short distance behavior of the potential.

Let us summarize our main findings \cite{smantovani}. We have established the ground for the description of  baryonic matter in the Chiral Soliton Model.  The main difference between this description and the Skyrme model one is the existence of quark fields in the Lagrangian. Color has been eliminated in favor of a scalar isoscalar field responsible for confinement. In this way the color phase transition and the chiral phase transition are governed by the same mechanism, namely the non topological structure of the $\sigma$  field soliton.

As a first step  we have recovered in our Chiral Soliton Model  the B=1 hedgehog solution. The solitonic structure is apparent in the solution leading to a  confinement scenario, where the pion field is partially expelled from the quarkish core. Thus the center of the baryons is dominated by relativistic quarks bouncing around, in the intermediate region quarks and pions coexist and in the outer region a pion cloud extends out  for quite some distance \cite{DV11a}. 

We have studied the B=2 interaction, with a product Ansatz approach, and discovered  the dominating features of the baryon-baryon interaction. By looking at the energy density as the baryons approach each other, we show that the long range tail of the baryon-baryon interaction is dominated by the pion cloud. We have analyzed the importance of  the quills orientation into the dynamics and have found two sensitive orientations, which should  describe the two states of matter as determined by the realization of chiral symmetry,  a scenario which resembles closely our Skyrme description \cite{Lee:2003aq}.



\end{document}